\documentclass[preprint,superscriptaddress,showpacs,preprintnumbers,amsmath,amssymb]{revtex4}

\usepackage{graphicx}
\usepackage{dcolumn}
\usepackage{bm}
\usepackage{color}

\begin{document}


\title{Deformation effect on the center-of-mass correction energy in nuclei ranging from Oxygen to
Calcium \footnote{Supported by the Major State Basic Research
Development Program (2007CB815000), and the National Natural Science
Foundation of China (10775004 and 10975008)}}

\author{ZHAO Peng-Wei}
 \affiliation{School of Physics and State Key Laboratory of Nuclear Physics and Technology, Peking University, Beijing 100871}

\author{SUN Bao-Yuan}%
 \affiliation{School of Physics and State Key Laboratory of Nuclear Physics and Technology, Peking University, Beijing 100871}

\author{MENG Jie \footnote{Email: mengj@pku.edu.cn}}
 \affiliation{School of Physics and Nuclear Energy Engineering, Beihang University, Beijing 100191, China}
 \affiliation{School of Physics and State Key Laboratory of Nuclear Physics and Technology, Peking University, Beijing 100871}
 \affiliation{Department of Physics, University of Stellenbosch, Stellenbosch, South Africa}%


\begin{abstract}
The microscopic center-of-mass (c.m.) correction energies for nuclei
ranging from Oxygen to Calcium are systematically calculated by both
spherical and axially deformed relativistic mean-field (RMF) models
with the effective interaction PK1. The microscopic c.m. correction
energies strongly depend on the isospin as well as deformation and
deviate from the phenomenological ones. The deformation effect is
discussed in detail by comparing the deformed with the spherical RMF
calculation. It is found that the direct and exchange terms of the
c.m. correction energies are strongly correlated with the density
distribution of nuclei and are suppressed in the deformed case.
\end{abstract}
\pacs{
21.60.Jz,  
21.10.Gv,  
24.10.Jv,  
27.30.+t,  
27.40.+z   
}

\maketitle


The mean field approximation is one of the most successful
theoretical approaches in quantitatively describing the properties
of both nuclear matter and finite nuclei near or far from the
stability line. However, for a finite nuclear system, the
translational symmetry of ground-state wave function is violated due
to the localization of the center-of-mass (c.m.) in the mean field
potential. In comparison with the preservation of rotational
symmetry for spherical nuclei and/or particle-number symmetry for
closed shell nuclei, the translational symmetry violation, probably
as the most important case of symmetry breaking, is compulsory for
all the nuclei. Therefore, it is necessary to develop proper methods
for the translational symmetry restoration.

A rigorous way to restore the broken translational symmetry is the
projection method, namely, projecting the ground-state wave function
onto a good c.m. momentum. In principle, Variation-after-projection
(VAP)~\cite{Schmid1987} is an ideal solution in comparison with
projection-after-variation (PAV) since it restores full Galilean
invariance~\cite{Ring1980B}. However, it is numerically too
expensive and impractical to be used in large-scale investigations.
Hence, PAV is often used as a simpler treatment to give the c.m.
correction energy. For the sake of feasibility and transferability,
a standard way, i.e., expanding the correction in orders of the
total momentum in c.m. frame $\langle\bm{P}^{2n}_{\rm c.m.}\rangle$
and stopping at first order, is suggested, which is denoted as
microscopic c.m. correction method~\cite{Bender2000}. Besides,
phenomenological c.m. correction is also widely used in practical
applications~\cite{Butler1984,Friedrich1986}. It has been shown that
the c.m. correction gives a remarkable contribution to the total
binding energy in light nuclei (e.g., about $9\%$ in
$^{16}$O)~\cite{Long2004}.

As one of the most successful representatives of mean field theory,
the relativistic mean-field (RMF) theory~\cite{Serot1986} has
received a great deal of attention during the past
decades~\cite{Ring1996,Meng2006}. In RMF theory, both the
phenomenological and microscopic c.m. correction are adopted to give
the c.m. correction energy. Therefore, it is interesting to
investigate the differences between these two c.m. correction
methods. Since the microscopic c.m. correction energy is decided by
the ground-state wave function, it is expected that it depends not
only on the mass number, but also on the deformation of nuclei.
While in the phenomenological case, the deformation effect usually
does not account for the c.m. correction energy. So far, a
systematic study of the deformation effect on the microscopic c.m.
correction energy in a large-scale nuclear mass region has not been
given.

In this letter, the microscopic c.m. correction energies for nuclei
ranging from Oxygen to Calcium are investigated systematically in
the spherical and axially deformed RMF models, and compared with the
phenomenological ones. Furthermore, the deformation effects on the
c.m. correction energies are studied in detail.


The starting point of the RMF theory is an effective Lagrangian
density where nucleons are described as Dirac spinors $\psi$ which
interact via the exchange of several mesons (the isoscalar scalar
$\sigma$, the isoscalar vector $\omega$, and isovector vector
$\rho$) and the photon~\cite{Serot1986}. The detailed formulation of
the RMF theory can be found in Ref.~\cite{Ring1996,Meng2006}.

The microscopic c.m. correction energy is given by
\begin{equation}\label{Ecmmicro}
    E_{\rm c.m.}^{\rm mic}=-\frac{1}{2MA}\left.\langle\bm{P}^2_{\rm
    c.m.}\rangle\right.,
\end{equation}
where $\bm{P}_{\rm c.m.}=\sum\limits_i^A\bm{p}_i$, which is
given by the sum of the single-particle momentum operators
$\bm{p}_i$, is the total momentum operator in the c.m. frame. The
expectation value of $\bm{P}_{\rm c.m.}^2$ is
\begin{equation}\label{Ecmmicrop2}
    \left.\langle
    \bm{P}^2_{\rm c.m.}\rangle\right.=\sum\limits_ap_{aa}^2-\sum\limits_{a,b}\bm{p}_{ab}\cdot\bm{p}_{ab}^\ast,
\end{equation}
 where $a$ and $b$ denote the occupied single-particle states.
 The expectation value of $\bm{p}_i^2$ in the state $|a\rangle$
 is denoted as $p_{aa}^2$, and $\bm{p}_{ab}$ is the
off-diagonal single-particle matrix element between the state
$|a\rangle$ and $|b\rangle$. Therefore, the correction energy in Eq.
(\ref{Ecmmicro}) can be decomposed into the direct term $E_{\rm
c.m.}^{\rm dir}$ and the exchange term $E_{\rm c.m.}^{\rm exc}$,
 \begin{subequations}
 \begin{eqnarray}\label{Ecmmicrodir}
    E_{\rm c.m.}^{\rm dir}&=&-\frac{1}{2MA}\sum\limits_ap_{aa}^2,\\
    \label{Ecmmicroexch}
    E_{\rm c.m.}^{\rm exc}&=&\frac{1}{2MA}\sum\limits_{a,b}\bm{p}_{ab}\cdot\bm{p}_{ab}^\ast .
 \end{eqnarray}
 \end{subequations}
It shows that $E_{\rm c.m.}^{\rm dir}$ increases while $E_{\rm
c.m.}^{\rm exc}$ decreases the binding energy of a given nuclei. The
further evaluations of Eq.~(\ref{Ecmmicrop2}) in spherical and
axially symmetry are outlined in Ref.~\cite{Bender2000}.

As the microscopic calculation of $E_{\rm c.m.}^{\rm mic}$ in Eq.
(\ref{Ecmmicro}) is often very time consuming, several
phenomenological approaches are proposed, including the
phenomenological formulas from harmonic oscillator states,
\begin{equation}\label{Ecmosc}
    E_{\rm c.m.}^{\rm osc} = -\frac{3}{4}41A^{-1/3}~{\rm MeV},
\end{equation}
and a fit to the microscopic c.m. correction energies calculated
with the Skyrme interaction $Z_\sigma$~\cite{Friedrich1986},
\begin{equation}\label{Ecmfit}
    E^{\rm fit}_{\rm c.m.} = -17.2 A^{-0.2}~{\rm MeV}.
\end{equation}

In present work, the microscopic c.m. correction energies for nuclei
with $8\leq Z\leq20$ are calculated in both the spherical and
axially deformed RMF theory with the non-linear effective
interaction PK1~\cite{Long2004}. In the calculation, the time-odd
component for odd-A and odd-odd nuclei~\cite{Yao06} is not included
as its influence on the c.m. correction energy is
negligible~\cite{Chen07}. The Dirac equation for nucleons and the
Klein-Gordon equations for mesons are solved using the expansion
method with the harmonic-oscillator basis~\cite{Gambhir1990}. In the
following investigation, 14 shells are used for both the fermion
fields and the meson fields. As the microscopic c.m. correction
energies are the main concern here, the pairing correlations are not
included.

\begin{figure*}[htbp!]
    \includegraphics[width=0.6\textwidth]{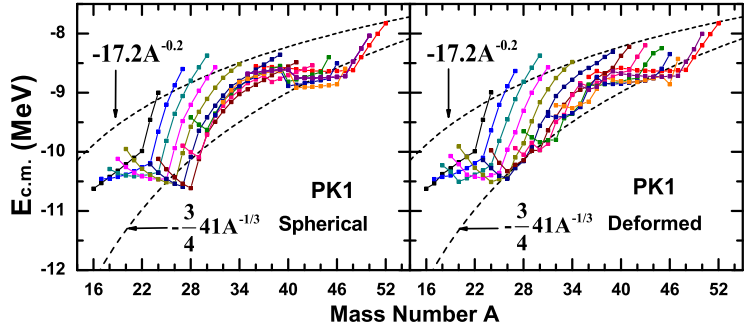}
    \caption{Microscopic c.m. correction energies $E_{\rm
c.m.}^{\rm mic}$ (solid lines) of nuclei with $8\leq Z\leq20$ in the
spherical (left panel) and axially deformed (right panel) RMF
calculations with the effective interaction PK1, in comparison with
two phenomenological results $E_{\rm c.m.}^{\rm osc}$ and $E^{\rm
fit}_{\rm c.m.}$ (dashed lines). The solid lines from the left to
the right respectively correspond to the isotopic chains from Oxygen
to Calcium.}
    \label{Fig1}
\end{figure*}

The microscopic c.m. correction energies $E_{\rm c.m.}^{\rm mic}$ of
the nuclei ranging from Oxygen to Calcium calculated in the
spherical and axially deformed RMF theory are shown in
Fig.~\ref{Fig1} as functions of the mass number $A$ and compared
with the phenomenological $E_{\rm c.m.}^{\rm osc}$ and $E^{\rm
fit}_{\rm c.m.}$. It is found that both the microscopic and
phenomenological c.m. correction energies increase with the mass
number systematically. $E^{\rm fit}_{\rm c.m.}$ is always larger
than $E_{\rm c.m.}^{\rm osc}$ in this mass region, and the
microscopic c.m. correction energies of most nuclei are in between
with strong isospin dependence. Generally speaking, $E_{\rm
c.m.}^{\rm fit}$ is more suitable for neutron-rich nuclei, whereas
$E_{\rm c.m.}^{\rm osc}$ for nuclei around $N=Z$.

\begin{figure}[htbp!]
    \includegraphics[width=0.3\textwidth]{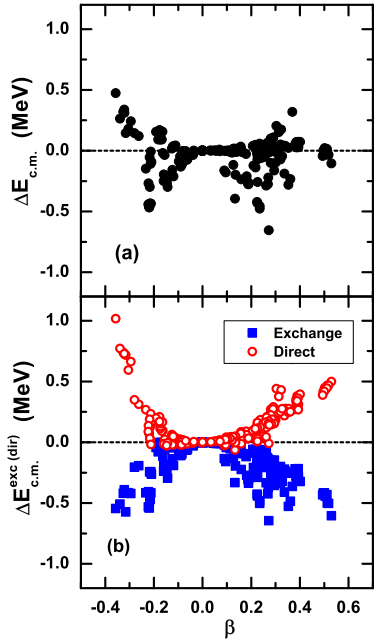}
    \caption{Differences of the microscopic c.m.
correction energy $\Delta E_{\rm c.m.}=E_{\rm c.m.}^{\rm def}-E_{\rm
c.m.}^{\rm sph}$ (upper panel) and their corresponding direct term
$\Delta E_{\rm c.m.}^{\rm dir}$ (open circles) and exchange term
$\Delta E_{\rm c.m.}^{\rm exc}$ (filled squares) (lower panel)
between deformed RMF calculations $E_{\rm c.m.}^{\rm def}$ and the
corresponding spherical ones $E_{\rm c.m.}^{\rm sph}$ for nuclei
with $8\leq Z\leq20$ as functions of the deformation parameter
$\beta$. }
    \label{Fig2}
\end{figure}

From Fig.~\ref{Fig1}, the deformation effects on the microscopic
c.m. correction energies are revealed by comparing the spherical and
deformed results. Such deformation effects are extracted from the
differences of microscopic c.m. correction energies between deformed
RMF calculations $E_{\rm c.m.}^{\rm def}$ and spherical ones $E_{\rm
c.m.}^{\rm sph}$, i.e., $\Delta E_{\rm c.m.}=E_{\rm c.m.}^{\rm
def}-E_{\rm c.m.}^{\rm sph}$, and illustrated in Fig.~\ref{Fig2}(a)
as a function of the quadrupole deformation parameter $\beta$
obtained in the  deformed RMF calculations. For $|\beta|<0.1$,
$\Delta E_{\rm c.m.}$ almost vanishes. While, for $|\beta|>0.1$,
most of the $| \Delta E_{\rm c.m.}|$ increase with $|\beta|$ upto
about $0.5$~MeV.

In order to understand the non-unilateral effect of deformation on
the microscopic c.m. correction energies, the direct  $E_{\rm
c.m.}^{\rm dir}$ and exchange term $E_{\rm c.m.}^{\rm exc}$ in
Eqs.~(\ref{Ecmmicrodir}) and ~(\ref{Ecmmicroexch}) are calculated,
respectively. Their corresponding differences $\Delta E_{\rm
c.m.}^{\rm dir}$ and $\Delta E_{\rm c.m.}^{\rm exc}$ between the
deformed and spherical calculations are shown in Fig.~\ref{Fig2}(b)
as functions of the quadrupole deformation parameter $\beta$.
Different from $\Delta E_{\rm c.m.}$, it is found that both $\Delta
E_{\rm c.m.}^{\rm dir}$ and $\Delta E_{\rm c.m.}^{\rm exc}$ vary
monotonously with $|\beta|$. Due to the different signs in $E_{\rm
c.m.}^{\rm dir}$ and $E_{\rm c.m.}^{\rm exc}$,  $\Delta E_{\rm
c.m.}^{\rm dir}$ increases with deformation up to $1$ MeV and
$\Delta E_{\rm c.m.}^{\rm exc}$ decreases with deformation down to
$-0.6$ MeV. Therefore, for a given nucleus, both spherical $|E_{\rm
c.m.}^{\rm dir}|$ and $|E_{\rm c.m.}^{\rm exc}|$ are found to be
larger than their corresponding deformed ones and the non-unilateral
effect of deformation on the microscopic c.m. correction energies is
just due to the competition between $E_{\rm c.m.}^{\rm dir}$ and
$E_{\rm c.m.}^{\rm exc}$.

\begin{figure}[htbp!]
    \includegraphics[width=0.3\textwidth]{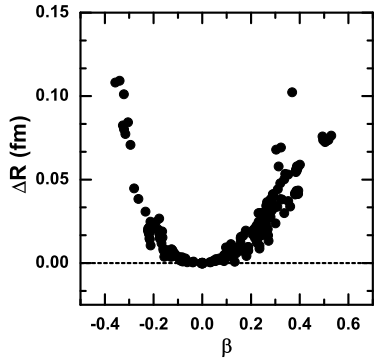}
    \caption{Differences of the matter rms-radii ($\Delta R=R_{\rm
def}-R_{\rm sph}$) between axially deformed RMF calculations $R_{\rm
def}$ and the corresponding spherical ones $R_{\rm sph}$ for nuclei
with $8\leq Z\leq20$ as a function of the deformation parameter
$\beta$.}
    \label{Fig3}
\end{figure}

Since the matter rms-radii as well as the microscopic c.m.
correction energies are measures for the localization of the
many-body wave function, it is interesting to investigate their
correlations. In Fig.~\ref{Fig3} are shown the differences of the
matter rms-radii (i.e., $\Delta R=R_{\rm def}-R_{\rm sph}$) between
$R_{\rm def}$ given by axially deformed RMF calculations and $R_{\rm
sph}$ by spherical ones as a function of the quadrupole deformation
parameter $\beta$. It is clear that $\Delta R$ increases
monotonously upto the maximum ($\thicksim0.1$ fm) with $|\beta|$,
and exhibits a similar behavior as $\Delta E_{\rm c.m.}^{\rm dir}$
and $|\Delta E_{\rm c.m.}^{\rm exc}|$ shown in Fig.~\ref{Fig2}(b).
In addition, $R_{\rm def}$ is always larger than $R_{\rm sph}$. As
larger radius corresponds to smaller $p^2_{aa}$ and
$\bm{p}_{ab}\cdot\bm{p}_{ab}^\ast$ in Eqs.~(\ref{Ecmmicrodir}) and
~(\ref{Ecmmicroexch}), it leads to a suppression on both the direct
and exchange term of $E_{\rm c.m.}^{\rm mic}$ in the deformed RMF
calculations. Therefore, the direct term and exchange term of
$E^{\rm mic}_{\rm c.m.}$ serve also as measures for the density
distribution of nuclei.

In summary, a systematic study of the microscopic c.m. correction
energies for nuclei with $8\leq Z\leq20$ is performed by the
spherical and deformed RMF models with the effective interaction
PK1. The microscopic c.m. correction energies, which are found in
between the phenomenological $E_{\rm c.m.}^{\rm fit}$ and $E_{\rm
c.m.}^{\rm osc}$, strongly depend on the isospin as well as the
deformation of nuclei. The deformation effect on $E^{\rm mic}_{\rm
c.m.}$ is clarified by comparing the deformed and spherical RMF
calculations. In comparison with the spherical calculations, a
suppression on both the direct and exchange term of $E^{\rm
mic}_{\rm c.m.}$, which even reach $1$ MeV for the former and $0.6$
MeV for the latter, is found in the deformed case. Moreover, it is
illustrated that the direct and exchange terms of the c.m.
correction energies are correlated with the density distribution of
nuclei.

\end{document}